\documentclass[twocolumn,aps,prl,showpacs,superscriptaddress]{revtex4}
\usepackage{color}
\usepackage{graphicx}
\usepackage{SIunits}
\usepackage{anysize}
\marginsize{0.765 in}{0.765 in}{0.4 in}{0.4 in}

\begin{document}

\title{Classical Analogue of Electromagnetically Induced Transparency with a Metal/Superconductor Hybrid Metamaterial}

\author{Cihan Kurter}
\affiliation{Center for Nanophysics and Advanced Materials,
Department of Physics, University of Maryland, College Park,
Maryland 20742-4111}

\author{Philippe Tassin}
\affiliation{Ames Laboratory-U.S. DOE and Department of Physics and
Astronomy, Iowa State University, Ames, IA 50011, USA}
\affiliation{Applied Physics Research Group, Vrije
Universiteit Brussel, Pleinlaan 2, B-1050 Brussel, Belgium}

\author{Lei Zhang}
\affiliation{Ames Laboratory-U.S. DOE and Department of Physics and
Astronomy, Iowa State University, Ames, IA 50011, USA}

\author{Thomas Koschny}
\affiliation{Ames Laboratory-U.S. DOE and Department of Physics and
Astronomy, Iowa State University, Ames, IA 50011, USA}

\author{Alexander P. Zhuravel}
\affiliation{B. Verkin Institute for Low Temperature Physics and
Engineering, National Academy of Sciences of Ukraine, 61103 Kharkov,
Ukraine}

\author{Alexey V. Ustinov}
\affiliation{Physikalisches Institut and DFG-Center for Functional
Nanostructures (CFN), Karlsruhe Institute of Technology, D-76128
Karlsruhe, Germany}

\author{Steven M. Anlage}
\affiliation{Center for Nanophysics and Advanced Materials,
Department of Physics, University of Maryland, College Park,
Maryland 20742-4111}

\author{Costas M. Soukoulis}
\affiliation{Ames Laboratory-U.S. DOE and Department of Physics and
Astronomy, Iowa State University, Ames, IA 50011, USA}
\affiliation{Dept.\ of Material Science and Technology, and
Institute of Electronic Structure and Lasers (IESL), FORTH,
University of Crete, 71110 Heraklion, Crete, Greece}
\date{\today}

\begin{abstract}

Metamaterials are engineered materials composed of small electrical
circuits producing novel interactions with electromagnetic waves.
Recently, a new class of metamaterials has been created to mimic the
behaviour of media displaying electromagnetically induced
transparency (EIT). Here we introduce a planar EIT metamaterial that
creates a very large loss contrast between the dark and radiative
resonators by employing a superconducting Nb film in the dark
element and a normal metal Au film in the radiative element. Below
the critical temperature of Nb, the resistance contrast established
opens up a transparency window along with a large enhancement in
group delay, enabling a significant slowdown of electromagnetic
waves. We further show that superconductivity allows for precise
control of EIT response through changes in the superfluid density.
Such tunable metamaterials may be useful for telecommunication
purposes because of their large phase shifts and delay-bandwidth
products.

\end{abstract}


\maketitle

Electromagnetically induced transparency (EIT) is a quantum mechanical effect observed in three-level atomic systems in which a beam of light can find its way through the medium with almost no absorption~\cite{Matsko,Fleischhauer05}. The typical configuration involves two atomic states which can be independently excited to the same final state, but have a forbidden transition between them. A probe beam tuned near one allowed transition will show a Lorentzian absorption profile.  In the presence of a coherent pump beam exciting the other allowed transition, there can be interference that produces a narrow transparency window for the probe beam, with simultaneously strong dispersion resulting in a significant enhancement of the group delay and the possibility of slowdown/storage of light~\cite{Hau,Fleischhauer00,CLiu,Bajcsy}.

The characteristics of EIT can be reproduced by a classical resonator model based on two oscillating masses coupled by a spring~\cite{Alzar}. One mass with dissipation factor $\gamma_1$ is acted upon by an external force, whereas the other mass with dissipation of $\gamma_2$ is not driven directly. If $\gamma_2 << \gamma_1$, a greatly minimized absorption and steep dispersion are visible at a particular driving frequency, due to a superposition of normal modes with almost zero displacement of the lossy oscillator.

Several groups have recently demonstrated classical analogues of EIT, mostly with electromagnetic metamaterials made of resonant artificial elements~\cite{Fedetov,Papasimakis08,Tassin,Papasimakis09}. One approach is to create an array of rings with asymmetric splits enabling the excitation of trapped-current modes~\cite{Fedetov}. Another study has utilized a continuous fish-scale pattern to excite those normally inaccessible modes~\cite{Papasimakis08}. Highly symmetrical geometries have also been proposed, e.g., an array of double ring resonators for which the response does not depend on the polarization~\cite{Papasimakis09} and two mirrored split-ring resonators (SRRs) with a cut wire at the center achieving enhanced transmission without breaking the symmetry~\cite{LZhang}. Apart from metamaterials showing EIT at microwave frequencies, plasmonic structures have been designed~\cite{SZhang} to create similar characteristics at optical frequencies~\cite{NLiu10}.

Earlier efforts to demonstrate classical EIT using normal-metal
electromagnetic oscillators have suffered from insufficient loss
contrast between the dark and radiative resonators. Here, we present
a design with a resistance contrast that is several orders of
magnitude larger than previously reported. The proposed artificial
molecule (metamolecule) of our metal/superconductor hybrid
metamaterial consists of a double planar SRR made of a
superconducting Nb film symmetrically located around a normal-metal
(Au) strip (see the Supporting Material for the details of the
design and fabrication). Use of Nb limits operation to the sub-THz
band; using a high-$T_c$ superconductor instead of Nb might allow
scaling of our structure to THz frequencies. The Au strip is
oriented along the electric field of the fundamental waveguide mode
and is designed to have an electric dipole resonance around $f$ =
10.6 GHz; it is thus the radiative element in the metamolecule. Each
of the two SRRs have in general an electric dipole resonance
(currents in both arms flowing in the same direction) and a magnetic
dipole resonance (currents in both arms flowing in opposite
directions, i.e., current circling around the SRR)~\cite{NotePeaks}.
The position of the gaps in the SRRs has been chosen such that the
electric field of the fundamental waveguide mode cannot excite their
magnetic dipole resonance. In a free space experiment, the gaps
would have been in the middle of the SRRs, but here they are shifted
towards the center in order to account for the mode profile of the
incident wave. In this way, the magnetic dipole mode of the SRRs can
be used as the dark mode in our metamaterial.

Furthermore, the geometry has been optimized to increase the loss
contrast between the radiative (strip) and dark (SRRs) resonators.
Firstly, the width of the strip is small to increase its resistance.
Secondly, the mirror symmetry of the double SRR avoids magnetic
dipole radiation of the dark element~\cite{LZhang}. In order to
enhance the loss contrast, we have made the SRRs from a Nb thin
film. Below the $T_\mathrm{c}$ of Nb, the microwave surface
resistance of the dark element will be very small (about
\unit{20}{\micro\ohm} at \unit{10}{\giga\hertz} and
\unit{4}{\kelvin})~\cite{Pambianchi94}. In contrast, the strip is
made of a normal metal that retains significant surface resistance
at low temperatures.

We have performed the microwave transmittance and reflectance
measurements on this metamaterial by placing the metamolecule in a
Nb X-band waveguide. We have cooled the system down below the
$T_\mathrm{c}$ of Nb, and measured the scattering parameters. In the
transmittance spectrum in Fig.~1(a), we observe a broadband dip with
three narrow resonance-shaped features with increased transmittance.
This is quite different from earlier
works~\cite{Fedetov,Papasimakis08,Tassin,Papasimakis09,LZhang} where
only one narrow transmission peak is observed. The reflectance data
give sharp dips in each transparency window where the transmittance
is enhanced. The underlying Lorentzian stopband [dashed curve in
Fig.~1(b)] is due to the electric dipole interaction between the
wire and the incident field. It can also be shown from a simple
coupled-resonator model~\cite{Alzar,Tassin} that the
resonance-shaped features emerge when a dark resonator is coupled
very weakly to the radiative resonator and when the dark resonator
is of sufficiently high quality. The question that remains to be
addressed is why we find three of these features in our experiment.

\begin{figure}
\centering
\includegraphics[clip]{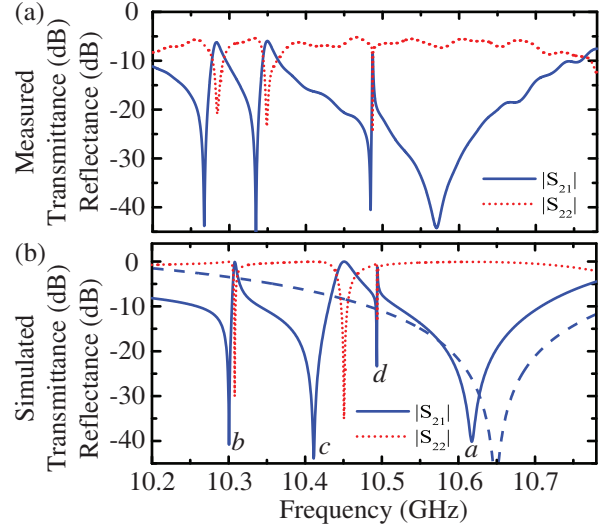}
\caption{(Color online) Transmittance and reflectance spectra. The transmittance shows a broad stopband originating
from the coupling of the electromagnetic waves to the electric
dipole mode of the wire, and three narrower resonance-like features
that are associated with dark modes of the SRRs. (a) Experimental
data at $T$ = \unit{2.6}{\kelvin}. The FWHM bandwidth of the resonant features is 9, 10, and 1 MHz for the transmittance peaks and 2, 1, and 0.1 MHz for the reflectance features. (b) Simulation data when the metamolecule is
horizontally displaced by \unit{0.1}{\milli\meter} from the center
of the waveguide. The dashed blue line is a simulation of the wire
only. The labels \textit{a}-\textit{d} correspond to the current
distributions in Fig.~2.}
\end{figure}

Our simulations (see the Supporting Material for details) can reproduce the experimental transmittance
spectrum to a high degree of similarity [Fig.~1(b)]. We find that
there is only one EIT-like transmission window if the sample is
positioned with perfect symmetry in the waveguide, and that the two
additional EIT features appear as soon as this symmetry is slightly
broken, i.e., by slightly tilting or shifting the metamolecule, or even by using a mesh that lacks the
required symmetry. Such deviations from the ideal position will
undoubtedly be present in the experiment. This suggests that there
are two extra dark modes that are uncoupled to
the wire if symmetrically located, but having a nonvanishing
coupling as soon as the symmetry is broken.

Further evidence for this hypothesis can be gained from the current
distributions obtained from the simulations shown in Figs.~2(a)-(d).
In Fig.~2(a), displaying the current distribution at $f$ =
\unit{10.62}{\giga\hertz} [the feature labeled \textit{a} in
Fig.~1(b)], there is a large current density in the wire only; this
is the electric dipole resonance of the wire. At the EIT feature at
$f$ = \unit{10.30}{\giga\hertz} [\textit{b} in Fig.~1(b)], we see
large currents in the SRRs and only a little current in the wire
commensurate with a dark mode excitation [Fig.~2(b)]. The current
distribution reveals that this dark mode is one of the
hybridizations of the electric dipole resonances of the SRRs. We see
that the currents in the two different SRRs flow in opposite
directions. Therefore, if the metamolecule is perfectly at the
center of the waveguide, there will be no coupling between the wire
and this dark mode and, hence, no EIT effect. However, if the
symmetry is slightly broken, the currents in the two SRRs are no
longer exactly opposite and a small coupling may be achieved. The
EIT feature at $f$ = \unit{10.41}{\giga\hertz} [\textit{c} in
Fig.~1(b)] has the main currents circling around the SRRs---one in a
clockwise direction, the other in a counterclockwise direction
[Fig.~2(c)]; this is the symmetric hybridization of the magnetic
dipole resonances of the SRRs that was observed in earlier
work~\cite{LZhang}. At the frequencies inside the EIT feature at $f$
= \unit{10.49}{\giga\hertz} [\textit{d} in Fig.~1(b)], the current
density is again circling the SRRs, but now in the same direction
for both SRRs [Fig.~2(d)]; this is the antisymmetric hybridization
of the magnetic dipole resonances of the SRRs. It may again be
inferred from symmetry considerations that this hybridization has
vanishing coupling to the electric dipole mode of the strip only if
the molecule is positioned exactly in the middle of the waveguide.
These considerations explain the origin of the three EIT features;
the resonance shape of the features is explained by constructive
(destructive) interference of the current induced in the wire by the
SRRs and the current directly excited by the external wave for the
peak (dip) in transmittance.

\begin{figure}
\centering
\includegraphics[clip]{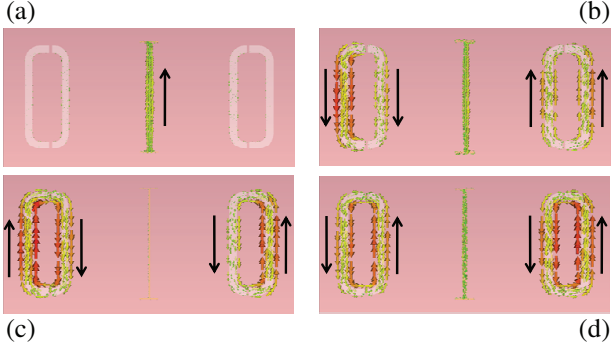}
\caption{(Color online) Current distributions in the wire and the
split-ring resonators. (a)~The electric dipole resonance of the
wire. (b)~The antisymmetric hybridization of the electric dipole
resonance of the SRRs. (c)~The symmetric hybridization of the
magnetic dipole resonance of the SRRs. (d)~The antisymmetric
hybridization of the magnetic dipole resonance of the SRRs.}
\end{figure}

We now proceed with our measurements of the group delay
[Fig.~3(a)]~\cite{Note}, which are reproduced well by the
simulations [Fig.~3(b)]. We observe significantly enhanced group
delay inside the three EIT transmission windows, in particular for
the highest-frequency EIT feature for which we find a group delay
above \unit{300}{\nano\second}. Indeed, the group delay enhancement
is related to the linewidth of the EIT feature, which is in turn
determined by the coupling strength~\cite{TassinOpt}. Since the
coupling of the highest-frequency dark mode is introduced by a
slight misalignment, it can be made arbitrarily small, unlike the
middle EIT feature for which the maximal group delay is limited by
the overall size of the metamolecule. With the use of a
superconductor, we can now achieve large group delays from a
subwavelength thin metamolecule, which may find application as
compact delay lines with phase shifts on the order
$\unit{10^4}{\radian}$. The delay-bandwidth product (DBP) of our
structure is 0.3, which approaches the DBPs in resonant cavities
(DBP $\approx 1$), but is still smaller than media containing atomic
vapors (DBP $\approx 10$) and some photonic crystal waveguides that
can achieve a DBP over 100, but we should keep in mind that the
latter systems are much longer than a single wavelength~\cite{Baba}.

\begin{figure}
\centering
\includegraphics[clip]{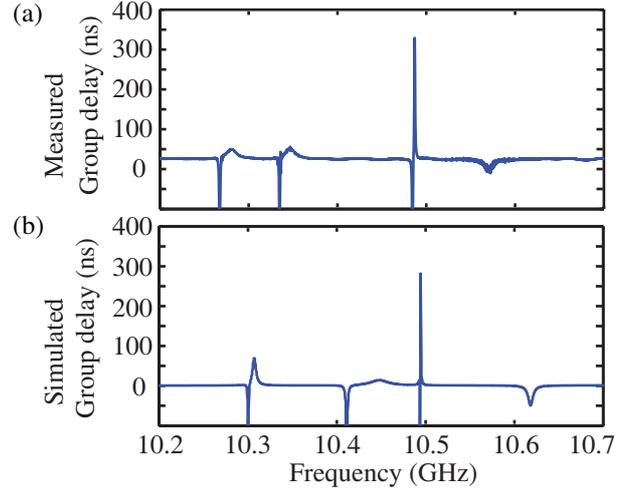}
\caption{(Color online) The group delay of the metamaterial obtained
(a)~from the experiment at $T$ = \unit{2.6}{\kelvin} (b)~from the
numerical simulations.}
\end{figure}

The reader should also note the transmittance contrast of about
\unit{40}{\deci\bel} between the peaks of the EIT windows and the
transmittance dip associated with the wire response. This has to be
compared with less than 10 dB for normal-metal EIT metamaterials
with similar geometry~\cite{LZhang}, i.e., a difference of three
orders of magnitude. The significantly improved transmittance
contrast is the direct consequence of the reduced loss in the dark
element by employing a superconductor. The presence of the
superconductor is critical in this experiment because the
weakly-coupled EIT resonances with large group delay are very
sensitive to loss and would be destroyed if a superconductor were
not used.

We investigate now the temperature dependence of the metamaterial
response. We have plotted the measured transmittance [Fig.~4(a)] and
group delay [Fig.~4(b)] for a set of temperatures. The enhancement
of the transmittance and the group delay [inset of Fig.~4(b)] both
decrease with increasing temperature. This is due to the increase in
ohmic loss of the superconductor, originating from a larger fraction
of electrons in the normal state, which at high temperatures finally
leads to an insufficient loss contrast to support classical EIT. We
also observe that the frequency of the EIT features shifts downward
with increasing temperature [inset of Fig.~4(a)], which is mainly
due to the increase in kinetic inductance of Nb as the superfluid
density decreases~\cite{Ricci06,Chen}. The EIT feature also broadens
with increasing temperature due to the increase in losses. The EIT
effect finally disappears at the critical temperature as Nb enters
the normal state and its ohmic losses exceed those of the Au
radiative resonator. The magnetic field required for tuning in this
structure is estimated to be on the order of
\unit{1}{\ampere\per\meter}. The switching time will likely be
limited by the time it takes to drive the superconducting resonators
into the normal state. We estimate this to be on the order of
microseconds for microwave magnetic fields. Other methods to tune
metamaterials at these frequencies include varactor diodes, MEMS
switches, photo-carrier generation, or mechanical tuning, all of
which involve additional losses, and none of them will be
significantly faster than our method.

\begin{figure}
\centering
\includegraphics[clip]{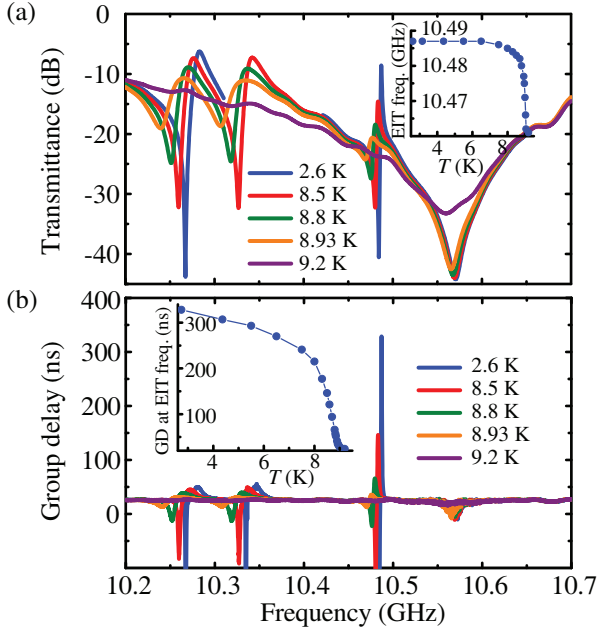}
\caption{(Color online) Temperature dependence of the measured
(a)~transmittance and (b)~group delay of the metamaterial. The
pronounced EIT features and the enhanced group delay at the lowest
temperature vanish if the temperature is increased above the
critical temperature of Nb. The inset in (a) shows the frequency of
the rightmost EIT feature as a function of temperature. The inset in
(b) shows the peak group delay at the rightmost EIT feature as a
function of temperature. The incident power was $-20$~dBm.}
\end{figure}

In summary, we have created a planar metal/superconductor hybrid
metamolecule that displays classical EIT with far greater loss
contrast, transmittance, and group delay than ever before
demonstrated. Under these circumstances, two new dark modes become
visible, and they can be designed to create multiple strong EIT
windows. By manipulation of the superconducting properties of the
dark resonators through temperature or magnetic field, the EIT
effects are tunable to an unprecedented extent.

The work at Maryland was supported by ONR Award No.\ N000140811058 and 20101144225000, the U.S.\ DOE (High Energy Physics) under Contract No.\ DESC0004950, the ONR/UMD AppEl Center, task D10 (Award No.\ N000140911190), and CNAM.  The work at Ames Lab was partially supported by the U.S.\ DOE (Basic Energy Science) under Contract No.\ DE-AC02-07CH11358 (computational studies), by ONR Award No.\ N000141010925 (characterization), and by the EU FET project PHOME, Contract No.\ 213310 (theory). The work in Karls\-ruhe is supported by the Fundamental Researches State Fund of Ukraine and the German International Bureau of the Federal Ministry of Education and Research (BMBF) under Grant Project No.\ UKR08/011, and a NASU program on ``nanostructures, materials and technologies.'' P.T. acknowledges a fellowship from the Belgian American Educational Foundation.


\begin{thebibliography}{22}
\bibitem{NotePeaks} The frequencies of the EIT resonances, i.e., of the dark modes can be positioned below or above the radiative resonance by proper design of the SRRs.
\bibitem{Note} The offset in the measured group delay is
caused by the delay of the waves in the transmission lines (coaxial
cables + waveguide) used in our measurements. We estimate this delay
to be \unit{26}{\nano\second}, consistent with the measured
background delay.
\end{thebibliography}

\end{document}